\documentclass[11pt,a4paper,onecolumn]{article}
\usepackage[left=1.6cm,right=1.6cm,top=1.6cm,bottom=2cm]{geometry}

\usepackage{amsmath,amssymb,amsfonts}
\usepackage{amsthm}
\usepackage{algorithmic}
\usepackage{graphicx}
\graphicspath{{numerics/figure/}}
\usepackage[font=footnotesize]{subfig}
\usepackage{textcomp}
\usepackage{array}
\usepackage{booktabs}
\usepackage{tabularx}
\usepackage{multirow}
\usepackage{color}
\usepackage{bm}
\usepackage{natbib}
\usepackage{float}

\newcommand{\mA}{{\bm{A}}}
\newcommand{\mL}{{\bm{L}}}
\newcommand{\mV}{{\bm{V}}}

\newcommand{\vone}{{\bm{1}}}
\newcommand{\va}{{\bm{a}}}
\newcommand{\vb}{{\bm{b}}}
\newcommand{\ve}{{\bm{e}}}
\newcommand{\vf}{{\bm{f}}}
\newcommand{\vg}{{\bm{g}}}
\newcommand{\vh}{{\bm{h}}}
\newcommand{\vs}{{\bm{s}}}
\newcommand{\vu}{{\bm{u}}}
\newcommand{\vv}{{\bm{v}}}
\newcommand{\vw}{{\bm{w}}}
\newcommand{\vx}{{\bm{x}}}
\newcommand{\vy}{{\bm{y}}}
\newcommand{\gE}{{\mathcal{E}}}
\newcommand{\gG}{{\mathcal{G}}}

\newcommand{\gV}{{\mathcal{V}}}
\newcommand{\sC}{{\mathbb{C}}}
\newcommand{\Tr}{\mathrm{Tr}}
\newcommand{\id}{\mathrm{id}}

\newcommand{\cL}{\mathcal{L}}
\newcommand{\cD}{\mathcal{D}}
\newcommand{\cE}{\mathcal{E}}
\newcommand{\cN}{\mathcal{N}}
\newcommand{\cK}{\mathcal{K}}
\newcommand{\cV}{\mathcal{V}}
\newcommand{\cB}{\mathcal{B}}
\newcommand{\cP}{\mathcal{P}}
\newcommand{\cC}{\mathcal{C}}

\newcommand{\QL}{\mathbb{L}}
\newcommand{\ket}[1]{|#1\rangle}
\newcommand{\bra}[1]{\langle #1|}
\newcommand{\proj}[1]{|#1\rangle\langle #1|}
\newcommand{\ketbra}[2]{|#1\rangle\langle #2|}
\newcommand{\vect}{\mathrm{vec}}

\newcommand{\At}{\widetilde{\mA}}
\newcommand{\xt}{\widetilde{\vx}}
\newcommand{\at}{\widetilde{\va}}

\newcommand{\rss}{\varrho_{\mathrm{ss}}}
\newcommand{\one}{\vone}
\newcommand{\lams}{\lambda_\star}
\newcommand{\lamb}{\bar{\lambda}}
\newcommand{\PP}{\cP^{\mathrm P}}
\newcommand{\PC}{\cP^{\mathrm C}}

\newtheorem{theorem}{Theorem}
\newtheorem{proposition}{Proposition}
\newtheorem{lemma}{Lemma}

\begin{document}

\title{\huge Quantum Network Flows That  Solve Linear Equations}

\author{
Chengyang~Jiang,
Zihao~Ren,
Lei~Wang,
and~Guodong~Shi%
\thanks{C. Jiang is with the School of Automation and Intelligent Manufacturing, Southern University of Science and Technology, Shenzhen 518055, China (e-mail: 12433001@mail.sustech.edu.cn).}
\thanks{Z. Ren and L. Wang are with the College of Control Science and Engineering, Zhejiang University, Hangzhou 310027, China (e-mail: zhren2000@zju.edu.cn; lei.wangzju@zju.edu.cn).}
\thanks{G. Shi is with the Australian Center for Robotics, School of Aerospace, Mechanical and Mechatronic Engineering, The University of Sydney, Sydney, NSW 2006, Australia (e-mail: guodong.shi@sydney.edu.au).}
}

\date{}

\maketitle

\begin{abstract}
We develop a distributed quantum network flow for solving linear
equations from locally stored row data. Local pumping enforces
each node's equation constraint, while neighboring registers
interact through swaps. The node marginals obey an exact, closed
flow combining consensus with local constraint enforcement.
For consistent equations, all node states converge exponentially
to a common state supported on the solution space; a unique
solution yields full-network convergence to the product solution state.
For inconsistent equations of full column rank, scaled encoding
and readout recover the least-squares solution through successive
large-gain and small-scale limits.
For consistent systems, the marginal convergence rate is
nondecreasing and concave in the swap gain, approaching a
graph-independent limit for fixed local dissipators.
Its leading finite-gain correction couples graph topology and
row placement through the Laplacian pseudoinverse, enabling
semidefinite optimization of edge rates under a fixed total
swap-rate budget.
For invertible systems, a dissipative HHL-type construction using
the same pumping law provides a comparison of solution-state
encodings, decay rates, and row-update locality under
squared-row-norm pumping. Numerical examples are provided to validate the presented results. 
\end{abstract}


\section{Introduction}
\label{sec:introduction}

Linear equations are a basic computational component
in estimation, optimization, and the numerical treatment of physical
models. When equation rows are stored at different locations,
their solution becomes a network computation problem: each node
has access to part of the data and cooperates with its neighbors
to recover a common solution
\cite{WangMouLianRen2019}.
Distributed linear-equation algorithms address this setting through
local updates and neighbor communication
\cite{MLM2015,Anderson2016,Liu2018}.
Dynamical approaches encode solutions as attractive equilibria
\cite{Brockett1991,HelmkeMoore1994}.
In particular, the consensus $+$ projection flow combines agreement
among nodes with enforcement of their local row constraints,
providing a continuous-time framework for consistent equations
and least-squares approximation \cite{SAH2017}.
For quantum states, the dynamics must preserve positivity and trace
while allowing recovery of the solution.

Quantum linear-system algorithms approach the same algebraic task
through quantum state preparation. The Harrow--Hassidim--Lloyd
(HHL) algorithm prepares a state proportional to the solution
of an invertible linear system \cite{HHL2009}.
Subsequent developments improve the dependence on precision
and conditioning through inverse approximation, eigenstate
filtering, and adiabatic evolution
\cite{CKS2017,LinTong2020,Costa2022}.
Variational quantum linear solvers use parameterized circuits
and classical optimization to prepare the solution state
\cite{BravoPrieto2023}.
These approaches form a broad range of quantum methods for
linear algebra, with applications in differential equations
and quantum machine learning \cite{Morales2025}.
They also make the representation of the data and the recovery
of solution information central parts of the solver design.

Networks of quantum processors introduce a further design dimension:
how computation is organized across local registers and their
interactions. Quantum-network architectures and distributed
computing protocols provide ways to coordinate such registers
\cite{Kimble2008,Wehner2018,Barral2025}.
For linear equations, Shen et~al. develop a distributed
variational solver that assigns matrix blocks to quantum processors
and couples local routines through classical
optimization and communication \cite{Shen2026}.

Open quantum systems offer a dynamical route to this design.
Lindblad generators describe Markovian quantum evolution and
provide a framework for engineering attractive states and
subspaces \cite{AltafiniTicozzi2012,Manzano2020}.
Controlled dissipation can stabilize prescribed quantum states
and support quantum computation
\cite{TicozziViola2008,Kraus2008,VWC2009}.
Shang constructs a single-register dissipative solver using an
augmented residual map, with jumps that reset residual population
to a common reference level \cite{Shang2026}.
On a quantum network, pairwise swaps generate gossip and
consensus processes that establish permutation symmetry and
agreement of node marginals \cite{MST2015,SDPJ2016}.
Their graph structure also supports spectral rate optimization
\cite{Jafarizadeh2017TAC}.

We construct a distributed quantum solver in which each local
pump transfers population into its own row-constraint subspace
and neighboring swaps directly couple the registers.
This mechanism differs from resetting residual population to a
common reference level or coordinating local variational routines
through classical optimization.
The resulting autonomous Lindblad dynamics admit an exact, closed
marginal flow combining quantum consensus with local constraint
enforcement. This structure links the physical network evolution
to solution recovery, including convergence to a solution subspace
when the equation has multiple solutions.

The network architecture also makes the interaction between
equation rows and graph topology a central design variable.
Our analysis identifies a graph-independent rate limit and a
Laplacian-pseudoinverse correction that quantifies the combined
effect of topology and row placement. These results turn the
convergence analysis into edge-rate design under a fixed budget.
The contributions are threefold.

\begin{itemize}
\item We construct a distributed quantum solver whose node
marginals obey an exact, closed flow combining quantum consensus
with local constraint enforcement.
This structure yields exponential convergence to a common
solution-space state for consistent equations and full-network
convergence when the solution is unique.
Scaled encoding and readout recover the least-squares solution
of inconsistent full-column-rank systems through successive
large-gain and small-scale limits.

\item We characterize the convergence rate through its monotonicity
and concavity in the swap gain, graph-independent limit, and
large-gain correction coupling topology and row placement.
These results yield semidefinite edge-rate optimization under
a fixed total budget, with sensitivity and allocation optimality
conditions.

\item We apply the same pumping law to projected equation rows
to construct a single-register dissipative HHL-type solver.
Conditional measurement relates its normalized solution-state
target to the distributed encoding.
We compare the data requirements and spectral decay rates of
the two constructions under squared-row-norm pumping.
\end{itemize}

\textit{Notation:}
We use standard matrix and Dirac notation.
The symbols $(\cdot)^\dagger$ and $\Tr$ denote conjugate transpose
and trace; $I$, $\id$, and $\one$ denote the identity operator,
identity map, and all-ones vector.
The space of operators on $\sC^d$ is $\cB(\sC^d)$,
and $\succeq$ denotes the positive semidefinite order. We use $\|\cdot\|$ for the Euclidean vector norm,
the matrix operator norm, and the Hilbert--Schmidt induced
superoperator norm; \mbox{$\|\cdot\|_1$} and \mbox{$\|\cdot\|_2$} denote
the trace and Hilbert--Schmidt norms.
The inner product is $\langle X,Y\rangle=\Tr(X^\dagger Y)$,
extended componentwise to tuples, so that
$\|\boldsymbol X\|_2^2=\sum_k\|X_k\|_2^2$.
For a column vector $\vv$, $(\vv,\tau)$ appends the scalar $\tau$.
The symbols $B^+$, $\lambda_{\min}^{+}(B)$, and
$\sigma_{\min}(B)$ denote the Moore--Penrose inverse, smallest positive
eigenvalue, and smallest singular value, respectively.
We write superoperator action as $P(X)$ and column-wise
vectorization as $\vect(X)$.
Unless indicated otherwise, logarithms are natural.

\section{Problem Setup and Solver Construction}
\label{sec:setup}

\subsection{Equation Encoding and Readout}
\label{sec:classical}

We study the distributed linear equation $\mA\vx=\vb$
over a quantum network of $N\ge2$ nodes, where
$\mA\in\sC^{N\times m}$, $\vb\in\sC^N$,
and $\vx\in\sC^m$ contains $m\ge1$ unknowns.
Each node $k$ holds the local equation
$\va_k^\dagger \vx=b_k$, with $\va_k\in\sC^m$
and $b_k\in\sC$, and interacts only
with its neighbors.

To express the affine equations as homogeneous
constraints, we introduce the augmented matrix
$\At=[\mA,-\vb]\in\sC^{N\times d}$,
where $d=m+1$.
Since
\begin{equation}
\At(\vx,1)=\mA\vx-\vb,
\label{eq:homogenization}
\end{equation}
the equation $\mA\vx=\vb$ is equivalent to
$(\vx,1)\in\cN$, where
$\cN=\ker\At\subseteq\sC^d$.
For a consistent equation with a unique
solution $\vx^\star$, we have
$\cN=\mathrm{span}\{\xt\}$, where
\begin{equation*}
\xt=
\frac{(\vx^\star,1)}
{\sqrt{1+\|\vx^\star\|^2}}.
\end{equation*}
If multiple solutions exist, then
$\dim\cN\ge2$, whereas $\cN=\{0\}$
for an inconsistent equation with $\mA$
of full column rank.

To encode the homogeneous constraints induced
by \eqref{eq:homogenization} into quantum registers,
we associate each node with a register
whose Hilbert space is $\sC^d$.
Each register can be embedded in
$\lceil\log_2 d\rceil$ qubits.
We use the orthonormal computational basis
$\{\ket{\ell}\}_{\ell=1}^d$, where the first
$m$ states represent the unknown
coordinates and $\ket d$ the homogenizing
coordinate.
Accordingly, any $\vv\in\sC^d$
is represented as
$\ket{\vv}=\sum_{\ell=1}^d v_\ell\ket\ell$.
Assuming that all rows of $\At$ are nonzero,
we define the normalized row vectors by
\begin{equation*}
\bra{\at_k}
=
\frac{(\va_k^\dagger,-b_k)}
{\|(\va_k^\dagger,-b_k)\|}.
\end{equation*}

Each local equation is encoded by a rank-one
projector onto its normalized augmented row,
with the corresponding orthogonal complement
and common constraint subspace given by
\begin{equation}
\label{eq:row-projectors}
\Pi_k=\ket{\at_k}\bra{\at_k},\quad
\Pi_k^\perp=I-\Pi_k,\quad
\cN=\bigcap_{k=1}^N\ker\Pi_k.
\end{equation}
For a single-register state $\varrho$ on $\sC^d$,
the $k$th row constraint is satisfied when
\begin{equation}
\label{eq:row-constraint}
\Tr(\Pi_k\varrho)
=\bra{\at_k}\varrho\ket{\at_k}=0.
\end{equation}
Since $\varrho$ is positive semidefinite,
\eqref{eq:row-constraint} is equivalent to
$\varrho\ket{\at_k}=0$, which holds if and
only if $\varrho$ is supported on $\ker\Pi_k$.
Consequently, by \eqref{eq:row-projectors},
all row constraints are satisfied exactly when
$\varrho$ is supported on $\cN$.

The homogenizing coordinate provides a reference for
recovering the original variables. For a state $\varrho$
with $\bra d\varrho\ket d>0$, define the readout map for
$\ell=1,\ldots,m$ by
\begin{equation}
\hat{\vx}(\varrho)=
(\hat x_1(\varrho),\ldots,\hat x_m(\varrho))^\top,
\quad
\hat x_\ell(\varrho)=
\frac{\bra{\ell}\varrho\ket{d}}
{\bra{d}\varrho\ket{d}}.
\label{eq:solution-readout}
\end{equation}
If $\varrho$ is supported on $\cN$, then
$\At\varrho=0$ implies $\mA\hat{\vx}(\varrho)=\vb$.
In the unique-solution case, the only state supported
on $\cN$ is $\varrho=\proj{\xt}$, which yields
$\hat{\vx}(\varrho)=\vx^\star$.
For inconsistent equations with $\mA$ of full column rank, the recovery target is the unique least-squares solution.

\subsection{Quantum Consensus Dynamics}
\label{sec:consensus}

To couple the local quantum registers, we consider
a fixed, connected, undirected, weighted graph
$\gG=(\gV,\gE)$, where
$\gV=\{1,\ldots,N\}$.
Each edge $\{j,k\}\in\gE$ has a symmetric
weight $a_{jk}=a_{kj}>0$, with $a_{jk}=0$
otherwise.
Let $\mL$ denote the weighted graph Laplacian
and $\lambda_2>0$ its algebraic connectivity
\cite{Chung1997}.

To establish quantum consensus over the network,
we employ the swap-based dynamics of
\cite{SDPJ2016}.
The joint network state $\rho$ is a density
operator on $(\sC^d)^{\otimes N}$, and
$\rho_k=\Tr_{\bar k}(\rho)$ denotes the
node state at node $k$, obtained by tracing out
all other registers.
For each edge $\{j,k\}\in\gE$,
let $U_{jk}$ denote the unitary operator
that swaps registers $j$ and $k$.
The resulting quantum consensus dynamics are
\begin{equation}
\dot\rho
=
\sum_{\{j,k\}\in\gE}
a_{jk}\bigl(U_{jk}\rho U_{jk}-\rho\bigr).
\label{eq:qc}
\end{equation}
These interactions couple neighboring
registers while preserving the trace
and positivity of the network state.

To analyze the swap dynamics
\eqref{eq:qc}, we introduce the
quantum Laplacian through vectorization.
Since each $U_{jk}$ is a real symmetric
permutation unitary, we obtain
\begin{equation*}
\frac{d}{dt}\vect(\rho)
=
-\QL\vect(\rho),
\end{equation*}
where $\QL$ is the quantum Laplacian
governing the swap dynamics of the
joint network state, defined by
\begin{equation*}
\QL
=
\sum_{\{j,k\}\in\gE}
a_{jk}
\bigl(I_{d^{2N}}-U_{jk}\otimes U_{jk}\bigr).
\end{equation*}

Unlike the graph Laplacian $\mL$, which acts
on node-level variables, $\QL$ acts on
the vectorized joint network state.
The operator $\QL$ is symmetric and
positive semidefinite, with its kernel
consisting of permutation-invariant
operators \cite[Lemma~2]{SDPJ2016}.
Moreover, by Aldous' spectral-gap theorem
\cite{CLR2010}, its smallest positive
eigenvalue equals the algebraic
connectivity $\lambda_2$ of $\mL$.
Consequently, the swap dynamics converge
exponentially to the permutation average
of the initial network state.
Swap dynamics establish permutation symmetry, and row-local pumping
supplies the constraint enforcement used by the solver.

\subsection{Local Pumping Dissipation}
\label{sec:dissipation}

To enforce the encoded row constraints, we
introduce local pumping dissipators acting
on the quantum registers.
For each node $k$, let $\gamma_k>0$ denote
the local pumping rate and define the
Lindblad dissipator
$\cD_k:\cB(\sC^d)\to\cB(\sC^d)$ by
\begin{equation}
\cD_k(\varrho)
=
\gamma_k
\left[
\bra{\at_k}\varrho\ket{\at_k}
\frac{\Pi_k^\perp}{m}
-
\frac{1}{2}
\left(
\Pi_k\varrho+\varrho\Pi_k
\right)
\right].
\label{eq:rowdiss}
\end{equation}
A Lindblad representation uses the jump operators
\begin{equation*}
J_{k,\ell}
=
\sqrt{\frac{\gamma_k}{m}}
\ketbra{e_{k,\ell}}{\at_k},
\end{equation*}
where $\{\ket{e_{k,\ell}}\}_{\ell=1}^m$
is an orthonormal basis of $\ker\Pi_k$.

As in optical pumping \cite{Happer1972},
these jumps transfer population from the
forbidden direction $\ket{\at_k}$ to the
constraint subspace $\ker\Pi_k$.
The population along $\ket{\at_k}$
decays at rate $\gamma_k$, while coherences
between $\ket{\at_k}$ and $\ker\Pi_k$
decay at rate $\gamma_k/2$.
States supported on $\ker\Pi_k$ are dark
states and remain stationary under $\cD_k$
\cite{Kraus2008,VWC2009}.
Conversely, the forbidden population of a stationary
state must vanish; positivity then confines its support
to $\ker\Pi_k$.

To characterize the spectral properties of
the local dissipators, we consider the
real vector space $\cV_0$ of traceless
Hermitian operators on $\sC^d$.
For any $Z\in\cV_0$, define the linear map
$P_k:\cV_0\to\cV_0$ by $P_k(Z)=-\cD_k(Z)$.
The orthogonal projections onto the population and coherence
subspaces are given by
\begin{equation*}
\begin{aligned}
\PP_k(Z)
&=
\bra{\at_k}Z\ket{\at_k}
\left(\Pi_k-\frac{\Pi_k^\perp}{m}\right),\\
\PC_k(Z)
&=
\Pi_kZ\Pi_k^\perp+\Pi_k^\perp Z\Pi_k.
\end{aligned}
\end{equation*}
The first projection extracts the population
imbalance between the forbidden direction
and its orthogonal complement, whereas
the second extracts their mutual coherences.
By \eqref{eq:rowdiss}, $P_k$ admits the decomposition
\begin{equation}
P_k=\gamma_k
\bigl(\PP_k+\tfrac{1}{2}\PC_k\bigr).
\label{eq:Pk}
\end{equation}
Orthogonality of the projections yields
\begin{equation*}
\langle Z,P_k(Z)\rangle
=\gamma_k\left(
\|\PP_k(Z)\|_2^2+
\frac12\|\PC_k(Z)\|_2^2
\right).
\end{equation*}
Together with $\Tr(Z)=0$, this orthogonal
decomposition gives the spectral characterization
and dissipation bounds in the following lemma.

\begin{lemma}
\label{lem:row}
A state $\varrho$ is stationary under
$\cD_k$ exactly when it satisfies
the local row constraint
$\bra{\at_k}\varrho\ket{\at_k}=0$
in \eqref{eq:row-constraint}.
The operator $P_k$ is symmetric and positive semidefinite,
with spectrum contained in $\{0,\gamma_k/2,\gamma_k\}$.
For every $Z\in\cV_0$, its quadratic form satisfies
\begin{equation}
\gamma_k\|Z\ket{\at_k}\|^2
\le
\langle Z,P_k(Z)\rangle
\le
\gamma_k\|Z\|_2^2.
\label{eq:Pk-bounds}
\end{equation}
In particular, $P_k(Z)=0$ if and only if
$Z\ket{\at_k}=0$.
\end{lemma}

Lemma~\ref{lem:row} relates the local
dissipation to the corresponding row
constraint. To characterize these
constraints collectively, we introduce
the penalty operator $Q$ and its
quadratic form, for any $\vv\in\sC^d$:
\begin{equation}
\label{eq:penalty}
Q=\sum_{k=1}^N\gamma_k\Pi_k,
\qquad
\vv^\dagger Q\vv
=\sum_{k=1}^N\gamma_k|\bra{\at_k}\vv|^2.
\end{equation}
Since all $\gamma_k>0$, the quadratic
form vanishes exactly on the common
constraint subspace $\cN$ defined in
\eqref{eq:row-projectors}, yielding
$\cN=\ker Q$.
In particular, $Q$ is positive definite
when $\mA\vx=\vb$ is inconsistent and $\mA$
has full column rank.
We denote the smallest positive
eigenvalue of $Q$ by $\mu$, and
the maximum local pumping rate by $\gamma_{\max}$.

\subsection{Distributed Quantum Solver}
\label{sec:network}

To construct a distributed quantum solver,
we combine the swap-based consensus dynamics
\eqref{eq:qc} with the local pumping
dissipators \eqref{eq:rowdiss}.
The resulting network dynamics are
\begin{equation}
\dot\rho
=
K\sum_{\{j,k\}\in\gE}
a_{jk}\bigl(U_{jk}\rho U_{jk}-\rho\bigr)
+
\sum_{k=1}^N\cD_k^{(k)}(\rho),
\label{eq:network}
\end{equation}
where $K>0$ is the swap gain
and $\cD_k^{(k)}$ denotes the extension
of $\cD_k$ to the full network, acting
on register $k$ and as the identity
superoperator on the remaining registers.
We denote the generator of
\eqref{eq:network} by $\cL$.
Throughout the network analysis, the dynamics
are given by \eqref{eq:network}, with the graph,
row assignment, and local dissipators fixed
unless explicitly varied or otherwise stated.

To characterize the evolution of individual
node states, we take the partial trace
$\Tr_{\bar k}$ of \eqref{eq:network}.
For each edge $\{j,k\}\in\gE$,
the swap operation satisfies
$\Tr_{\bar k}(U_{jk}\rho U_{jk})=\rho_j$,
while swaps between other nodes leave
$\rho_k$ unchanged.
Since the local dissipative dynamics preserve
the trace, terms acting on registers other
than $k$ vanish under the partial trace,
leaving only $\cD_k(\rho_k)$.
Consequently, each node state satisfies
\begin{equation}
\dot\rho_k
=
K\sum_{j=1}^N
a_{jk}(\rho_j-\rho_k)
+
\cD_k(\rho_k).
\label{eq:reduced}
\end{equation}
These equations define the reduced dynamics
of the network, a closed linear system with $N(d^2-1)$ real
state coordinates.
It provides the operator representation used for convergence analysis
and swap-rate optimization.

The network is distributed in both data and interactions. Each node implements the local dissipator \eqref{eq:rowdiss} using only its row data, while neighboring registers interact through swap operations at rates $Ka_{jk}$, as in quantum gossip \cite{MST2015}. By Lemma~\ref{lem:row}, local pumping suppresses deviations from the row constraints, while swaps promote agreement among node states. The reduced dynamics \eqref{eq:reduced} therefore combine consensus with local constraint enforcement, analogous to classical distributed projection flows \cite{SAH2017}, with the affine constraints encoded through homogenization.

The objective is to recover solutions of $\mA\vx=\vb$ from the node states. For consistent equations, we seek convergence to a common state supported on $\cN$, whose readout \eqref{eq:solution-readout} yields a solution whenever it is well-defined. If the solution is unique, the target network state is $(\proj{\xt})^{\otimes N}$. For inconsistent equations with $\mA$ of full column rank, the objective is instead to recover an approximation to the least-squares solution.

\section{Convergence and Rate Analysis}
\label{sec:main}

\subsection{Convergence for Consistent Equations}

We first study the convergence of the
node-state dynamics for consistent
linear equations and characterize
their common limiting state.
By Lemma~\ref{lem:row}, any state
$\varrho_0$ supported on $\cN$ is a
common stationary state of the local
dissipators.
Define $X_k(t)=\rho_k(t)-\varrho_0\in\cV_0$
and $\boldsymbol X(t)=(X_1(t),\ldots,X_N(t))$.
Using the node-state dynamics
\eqref{eq:reduced} and the local
dissipation operators \eqref{eq:Pk},
we obtain
\begin{equation}
\dot{\boldsymbol X}(t)=-M_K(\boldsymbol X(t)),
\quad
M_K=K\mL\otimes\id
+\mathrm{diag}(P_1,\ldots,P_N).
\label{eq:reduced-operator}
\end{equation}
The map $M_K$ is symmetric and
positive semidefinite. Its kernel is the same for
every $K>0$.

To identify the common limiting
node state, define the average
initial node state $\bar\rho$ and
the node-state vector
$\boldsymbol{\rho}(t)$ by
\begin{equation*}
\bar\rho=\frac{1}{N}\sum_{k=1}^N\rho_k(0),
\quad
\boldsymbol{\rho}(t)
=(\rho_1(t),\ldots,\rho_N(t)).
\end{equation*}
Let $\Pi_\cN$ denote the orthogonal
projector onto $\cN$. The common
limiting state and the corresponding
node-state vector are given by
\begin{equation}
\rho_\infty
=\Pi_\cN\bar\rho\,\Pi_\cN
+\frac{\Tr(\Pi_\cN^\perp\bar\rho)}
{\dim\cN}\Pi_\cN,
\quad
\boldsymbol{\rho}_\infty
=\one\otimes\rho_\infty.
\label{eq:limit}
\end{equation}

The convergence rate is characterized
by the positive eigenvalues of $M_K$.
We define the following rates:
\begin{equation}
\lambda_\star(K)
=\lambda_{\min}^{+}(M_K),
\qquad
\bar\lambda
=\lambda_{\min}^{+}(\bar P),
\label{eq:rate-definitions}
\end{equation}
where $\bar P=1/N\sum_{k=1}^N P_k$
is the averaged local dissipation
operator on $\cV_0$.
These rates turn the pumping and swap-rate choices into explicit
guarantees on the physical evolution time required for state preparation.
Resource costs depend on encoding, implementation, and readout.

For a unique solution, the node and network fidelities are
\begin{equation*}
f_k(t)=\bra{\xt}\rho_k(t)\ket{\xt},\quad
F(t)=\bra{\xt}^{\otimes N}\rho(t)\ket{\xt}^{\otimes N}.
\end{equation*}
The spectral gap of the full-network generator $\cL$ is
\begin{equation*}
g(K)=\min_{\lambda\in\mathrm{spec}(\cL)\setminus\{0\}}
\bigl(-\mathrm{Re}\,\lambda\bigr).
\end{equation*}

\begin{theorem}
\label{thm:consistent}
Suppose that $\mA\vx=\vb$ is consistent.
All node states converge exponentially
to the common limiting state
$\rho_\infty$ in \eqref{eq:limit}, satisfying
\begin{equation}
\|\boldsymbol{\rho}(t)
-\boldsymbol{\rho}_\infty\|_2
\le e^{-\lambda_\star(K)t}
\|\boldsymbol{\rho}(0)
-\boldsymbol{\rho}_\infty\|_2.
\label{eq:marginal_decay}
\end{equation}
If the solution is unique, then
$\rho_\infty=\proj{\xt}$, and the
node and network fidelities satisfy
\begin{equation}
0\le 1-f_k(t)\le 1-F(t)
\le \sqrt{2}\,N e^{-\lambda_\star(K)t}.
\label{eq:fidelity}
\end{equation}
The full-network spectral gap $g(K)$ also satisfies
\begin{equation}
\frac{\lambda_\star(K)}{2}
\le g(K)\le\lambda_\star(K).
\label{eq:full-gap-bound}
\end{equation}
\end{theorem}

Solution recovery uses the readout \eqref{eq:solution-readout}.
For a unique solution, $\rho_\infty=\proj{\xt}$,
and the readout denominator satisfies
\begin{equation*}
\bra d\rho_\infty\ket d
=\frac{1}{1+\|\vx^\star\|^2}>0.
\end{equation*}
For multiple solutions, $\rho_\infty$
generally depends on the initial condition.
Convergence to the solution space yields a valid readout provided
that $\bra d\rho_\infty\ket d>0$.
In particular, initializing all nodes
at the maximally mixed state gives
\begin{equation*}
\rho_k(0)=\frac{I}{d}
\quad\Longrightarrow\quad
\rho_\infty=\frac{\Pi_\cN}{\dim\cN},
\end{equation*}
which ensures a well-defined readout
for consistent equations. The readout ratios
can be estimated through repeated state
preparations and measurements, with
accuracy depending on the denominator.

Beyond solution recovery, we quantify
the evolution time required to achieve
a prescribed accuracy. For a unique
solution, the node-state convergence
bound \eqref{eq:marginal_decay} and
standard norm inequalities yield the
node trace-distance bound. Similarly,
the network fidelity bound
\eqref{eq:fidelity}, together with
the Fuchs--van de Graaf inequality
\cite{FuchsvandeGraaf1999}, gives
\begin{equation}
\begin{aligned}
\frac{1}{2}\|\rho_k(t)-\proj{\xt}\|_1
&\le \left(\frac{dN}{2}\right)^{1/2}
e^{-\lambda_\star(K)t},\\
\frac{1}{2}\|\rho(t)-\proj{\xt}^{\otimes N}\|_1
&\le (\sqrt{2}N)^{1/2}
e^{-\lambda_\star(K)t/2}.
\end{aligned}
\label{eq:trace-distance}
\end{equation}

These bounds yield sufficient evolution
times to achieve a prescribed trace-distance
accuracy $\varepsilon\in(0,1)$.
Specifically, the node and network
trace distances are bounded by
$\varepsilon$ for
$t\ge t_{\rm node}(\varepsilon)$
and $t\ge t_{\rm net}(\varepsilon)$,
respectively, where
\begin{equation*}
\begin{aligned}
t_{\rm node}(\varepsilon)
&=\frac{\log(1/\varepsilon)
+\frac12\log(dN/2)}
{\lambda_\star(K)},\\
t_{\rm net}(\varepsilon)
&=\frac{2\log(1/\varepsilon)
+\log(\sqrt{2}N)}
{\lambda_\star(K)}.
\end{aligned}
\end{equation*}

\subsection{Rate Analysis for Consistent Equations}

We study how the swap gain $K$ and the communication
graph affect the convergence rate $\lambda_\star(K)$
in \eqref{eq:reduced-operator}, with the local
constraints and pumping rates fixed.
In particular, we characterize the large-gain limit
and the finite-gain rate deficit caused by graph
topology and local dissipation heterogeneity.

Let $\bar{\cE}$ denote the eigenspace of $\bar P$
associated with $\bar\lambda$ in
\eqref{eq:rate-definitions}.
For any $\bar Z\in\bar{\cE}$ with $\|\bar Z\|_2=1$,
define the local dissipation deviations and their
graph-dependent coefficient by
\begin{equation}
\begin{aligned}
h_k=P_k(\bar Z)-\bar P(\bar Z),\quad
\kappa_G(\bar Z)
=\frac{1}{N}\sum_{j,k=1}^N
(\mL^+)_{jk}\langle h_j,h_k\rangle.
\end{aligned}
\label{eq:kappa-definition}
\end{equation}
Here $h_k$ measures the deviation from the averaged
dissipation along the slow mode $\bar Z$.
We set
$\kappa_G=\max_{\bar Z}
\kappa_G(\bar Z)\ge0$,
which captures the combined effect of local dissipation
heterogeneity and graph topology through $\mL^+$.

\begin{theorem}
\label{thm:rates}
Suppose that $\mA\vx=\vb$ is consistent.
The convergence rate $\lambda_\star(K)$
is nondecreasing and concave in $K$, with
\begin{equation}
\frac{\bar\lambda}{2}
\min\left\{
1,\frac{K\lambda_2}{2\gamma_{\max}}
\right\}
\le\lambda_\star(K)\le\bar\lambda,
\label{eq:ratebounds}
\end{equation}
where $\bar\lambda$ is the limiting
rate as $K\to\infty$ and satisfies
\begin{equation}
\frac{\mu}{N(m+1)}
\le\bar\lambda\le\frac{\mu}{2N}.
\label{eq:limiting-rate-bound}
\end{equation}
For $K>\bar\lambda/\lambda_2$,
the rate deficit satisfies
\begin{equation}
0\le\bar\lambda-\lambda_\star(K)
\le\frac{\gamma_{\max}^2}{K\lambda_2-\bar\lambda}.
\label{eq:saturation}
\end{equation}
In the limit $K\to\infty$, the rate deficit admits the expansion
\begin{equation}
\bar\lambda-\lambda_\star(K)
=\frac{\kappa_G}{K}+O(K^{-2}).
\label{eq:kappa}
\end{equation}
\end{theorem}

The limiting rate depends on the local equation constraints and
dissipation rates, independently of the communication graph.
At large finite gain, $\kappa_G$ determines the leading rate deficit
through the Laplacian pseudoinverse and local dissipation deviations.
It therefore supplies an equation-dependent criterion for network design.

\subsection{Swap-Rate Design Under a Fixed Budget}
\label{sec:design}

For consistent equations, the preceding rate analysis considers fixed
edge weights and a common swap gain $K$.
We now optimize individual edge swap rates
under a total budget, while keeping the
local dissipators and row assignment fixed.
Spectral optimization is widely used in
the design of fast mixing and consensus
dynamics \cite{BoydDiaconisXiao2004}.
We optimize the convergence rate of the coupled swap and
local-dissipation dynamics under a fixed total swap-rate budget.

Let $\vw=(w_e)_{e\in\gE}\ge0$
denote the edge swap rates; zero rates correspond
to unused candidate edges.
The original network corresponds to
$w_e=Ka_{jk}>0$, for which $M_{\vw}=M_K$.
The associated weighted Laplacian
$\mL_{\vw}$ gives the reduced operator
in \eqref{eq:reduced-operator} as
\begin{equation}
    M_{\vw}
    =\mL_{\vw}\otimes\id
    +\mathrm{diag}(P_1,\ldots,P_N).
\label{eq:design-operator}
\end{equation}

To distinguish a trial direction from the evolving
state difference $\boldsymbol X(t)$, let
$\boldsymbol Z=(Z_1,\ldots,Z_N)$ range over all
unit-norm operator tuples in
$\cK_0^\perp\subseteq\cV_0^N$, where
$\cK_0=\ker M_K$ for $K>0$.
The convergence rate is the Rayleigh minimum
\begin{equation}
    \lams(\vw)
    =\min_{\boldsymbol Z}
    \langle\boldsymbol Z,M_{\vw}(\boldsymbol Z)\rangle.
\label{eq:design-rate}
\end{equation}
This definition extends
\eqref{eq:rate-definitions} to general
edge-rate allocations, yielding zero
when additional nondecaying modes arise.
For a fixed total swap-rate budget $W>0$,
let $\vw=(w_e)_{e\in\gE}$ range over all
nonnegative allocations whose edge rates sum to $W$.
The optimal convergence rate is defined by
\begin{equation}
    \lams^{\rm opt}(W)
    =\max_{\vw}\lams(\vw).
\label{eq:design-objective}
\end{equation}

The optimization problem \eqref{eq:design-objective}
admits a semidefinite programming formulation.
Choose a fixed orthonormal coordinate basis of $\cV_0^N$.
Let $[M_{\vw}]$ denote the matrix representation of
$M_{\vw}$ in that basis, and let the columns of $\mV$
be coordinate vectors forming an orthonormal basis of
$\cK_0^\perp$. Under the same budget constraints,
the problem becomes
\begin{equation}
\begin{aligned}
    \max_{\vw,\eta}\quad &\eta\\
    \text{subject to}\quad
    &\mV^\dagger[M_{\vw}]\mV\succeq\eta I.
\end{aligned}
\label{eq:design-sdp}
\end{equation}
The optimized rates specify the edge swap rates in \eqref{eq:network}.

To separate budget and allocation, write $\vw=W\vs$, where $s_e\ge0$ for every edge and $\sum_e s_e=1$. For the coefficient-based analysis, we restrict $\vs$ to allocation proportions whose positive-weight edges form a connected graph. Let $\mL_{\vs}$ be the Laplacian with weights $s_e$, so that $\mL_{\vw}=W\mL_{\vs}$. For each $\bar Z$,
let $\kappa_G(\vs,\bar Z)$
denote the quadratic expression in
\eqref{eq:kappa-definition} with
$\mL_{\vs}$ in place of $\mL$, and set
\begin{equation*}
    \kappa_G(\vs)
    =\max_{\bar Z}
      \kappa_G(\vs,\bar Z).
\end{equation*}

\begin{proposition}
\label{prop:design}
The semidefinite program \eqref{eq:design-sdp} has optimal value
$\lams^{\rm opt}(W)$, which is nondecreasing and concave in $W$.
As the total budget increases, the optimal rate approaches $\lamb$:
\begin{equation*}
    \lim_{W\to\infty}\lams^{\rm opt}(W)=\lamb.
\end{equation*}
Moreover, $\kappa_G(\vs)$ is convex
in $\vs$. For fixed $\vs$
and sufficiently large $W$, the rate deficit satisfies
\begin{equation*}
    \lamb-\lams(W\vs)
    =\frac{\kappa_G(\vs)}{W}
    +O(W^{-2}).
\end{equation*}
\end{proposition}

For a unit-norm minimizing operator tuple $\boldsymbol Z$ in \eqref{eq:design-rate}, define the edge sensitivities $g_e=\|Z_j-Z_k\|_2^2$ and the sensitivity vector $\vg=(g_e)_{e\in\gE}$.

Let $\bar Z_1,\ldots,\bar Z_q$ form an orthonormal basis of $\bar{\cE}$, where $q=\dim\bar{\cE}$. Following \eqref{eq:kappa-definition}, define the local deviations and node potentials by
\begin{equation}
    h_{a,k}=P_k(\bar Z_a)-\bar P(\bar Z_a),
    \quad
    \Phi_{a,k}
    =\sum_{j=1}^N(\mL_{\vs}^{+})_{kj}h_{a,j}.
\label{eq:design-potentials}
\end{equation}
The edge potential drops are $\Phi_{a,j}-\Phi_{a,k}$. To express the allocation optimality condition through averaged squared potential drops, we assume either $q=1$ or that $\kappa_G(\vs,\bar Z)$ is independent of $\bar Z$ for each admissible $\vs$.

\begin{theorem}
\label{thm:design}
Suppose that $\mA\vx=\vb$ is consistent. For two rate allocations $\vw$ and $\vw'$, the sensitivity vector $\vg$ computed at $\vw$ is a supergradient of $\lams$, satisfying
\begin{equation}
    \lams(\vw')-\lams(\vw)
    \le\sum_e g_e(w'_e-w_e).
\label{eq:design-supergradient}
\end{equation}
Under the stated assumption, an admissible
$\vs^\star$ minimizes
$\kappa_G(\vs)$ precisely when its
potentials satisfy the following inequality
on every edge, with equality on used edges:
\begin{equation}
    \frac1q\sum_{a=1}^q
    \|\Phi_{a,j}-\Phi_{a,k}\|_2^2
    \le N\kappa_G(\vs^\star).
\label{eq:equal-drop}
\end{equation}
\end{theorem}

Theorem~\ref{thm:design} provides two complementary design criteria. The supergradient measures edge-rate sensitivity and guides iterative budget redistribution, while the equal-drop condition verifies whether allocation proportions minimize $\kappa_G(\vs)$ under the stated assumption. The latter targets the leading large-budget rate deficit, whereas \eqref{eq:design-sdp} maximizes the actual convergence rate at any budget.

\subsection{Least-Squares Recovery for Inconsistent Equations}
\label{sec:inconsistent}

For inconsistent linear equations, ordinary least squares
(LS) provides an approximate solution by minimizing the
residual. Let $\mA\vx=\vb$ be inconsistent, with $\mA$ of full
column rank. The unique LS solution is
\begin{equation}
    \vx_{\rm LS}
    =
    \arg\min_{\vx\in\sC^m}\|\mA\vx-\vb\|^2
    =
    (\mA^\dagger \mA)^{-1}\mA^\dagger \vb.
\label{eq:ls}
\end{equation}
For inconsistent equations, the recovery target is the unique
least-squares solution $\vx_{\rm LS}$.
We realize this target through scaled row encoding and an associated readout.

To relate the homogenized penalty to the
residual, we choose the pumping rates as the squared
norms of the augmented rows. This cancels the row
normalization, yielding $Q=\At^\dagger\At$
in \eqref{eq:penalty}. The rates and Rayleigh quotient are
\begin{equation}
    \gamma_k=\|(\va_k^\dagger,-b_k)\|^2,
    \quad
    \frac{(\vx,1)^\dagger Q(\vx,1)}{\|(\vx,1)\|^2}
    =
    \frac{\|\mA\vx-\vb\|^2}{1+\|\vx\|^2}.
\label{eq:tls-objective}
\end{equation}
This is the unweighted total-least-squares (TLS)
objective, which allows corrections to both
$\mA$ and $\vb$. The denominator $1+\|\vx\|^2$
arises from normalizing the augmented vector
and distinguishes this quotient from the
LS objective in \eqref{eq:ls}.

This distinction motivates a rescaled encoding.
For an encoding scale $\beta\in(0,1]$, consider the scaled system
and pumping rates
\begin{equation*}
    \mA\vy=\beta \vb,\quad \vy=\beta \vx,\quad
    \gamma_k(\beta)=\|(\va_k^\dagger,-\beta b_k)\|^2.
\end{equation*}
Define the scaled augmented matrix
$\At(\beta)=[\mA,-\beta \vb]$ and the corresponding
penalty $Q(\beta)=\At(\beta)^\dagger\At(\beta)$.
The scaled rows and rates determine the local
dissipators $\cD_k^\beta$ through
\eqref{eq:rowdiss}, while the swap terms
remain unchanged. For fixed $\vx$, the scaled
Rayleigh quotient, divided by $\beta^2$,
satisfies
\begin{equation*}
    \frac{\|\mA\vx-\vb\|^2}{1+\beta^2\|\vx\|^2}
    \xrightarrow{\beta\to0}\|\mA\vx-\vb\|^2.
\end{equation*}
Since $\vx=\vy/\beta$, the readout
\eqref{eq:solution-readout} becomes
\begin{equation*}
    \hat x_\ell^\beta(\varrho)
    =\frac{1}{\beta}\hat x_\ell(\varrho)
    =\frac{\bra\ell\varrho\ket d}
          {\beta\bra d\varrho\ket d}.
\end{equation*}

Let $\rho_k^{K,\beta}$ denote the equilibrium
state at node $k$ under the reduced dynamics
\eqref{eq:reduced} with the scaled dissipators.
We also consider the averaged single-register
dynamics
\begin{equation}
    \dot\varrho
    =\frac{1}{N}\sum_{k=1}^N\cD_k^\beta(\varrho).
\label{eq:averaged-dynamics}
\end{equation}

\begin{theorem}
\label{thm:inconsistent}
Suppose that $\mA\vx=\vb$ is inconsistent with $\mA$ of
full column rank. Under the scaled encoding with squared-row-norm
pumping rates, each node state $\rho_k(t)$ converges exponentially
to a unique equilibrium $\rho_k^{K,\beta}$ for every fixed
$\beta\in(0,1]$.
The equilibria converge to the unique stationary state
$\rss(\beta)$ of \eqref{eq:averaged-dynamics}:
\begin{equation*}
    \lim_{K\to\infty}\rho_k^{K,\beta}
    =\rss(\beta).
\end{equation*}
For all sufficiently small $\beta>0$, the
scaled readout is well-defined
at $\rss(\beta)$ and satisfies
\begin{equation}
    \hat{\vx}^\beta(\rss(\beta))
    =\vx_{\rm LS}+O(\beta^2).
\label{eq:ls-limit}
\end{equation}
\end{theorem}

Consequently, for every node $k$, the successive limits yield
\begin{equation*}
    \lim_{\beta\to0}\lim_{K\to\infty}
    \hat{\vx}^\beta(\rho_k^{K,\beta})
    =\vx_{\rm LS}.
\end{equation*}
For a target accuracy, choose $\beta$ sufficiently small
to control the encoding bias, then choose $K$ sufficiently large
for this scale and evolve long enough to approximate
$\vx_{\rm LS}$.

The encoding scale $\beta$ controls both the $O(\beta^2)$
encoding bias and the precision needed for readout.
For each coordinate, an additive error $\delta$ in the numerator
produces a readout error of at most $\varepsilon$ if
\[
|\delta|\le\varepsilon\beta\bra d\varrho\ket d,
\]
with the denominator held fixed.
This error bound allows the encoding scale and measurement precision
to be selected jointly to control encoding bias and coordinatewise
measurement error.

\section{Dissipative HHL-Type Solver}
\label{sec:hhl}

\subsection{Pumping Construction and Convergence}
\label{sec:hhl-me}

We apply the pumping dissipator of the distributed quantum solver
to normalized solution-state preparation, the target of HHL and
related quantum linear-system algorithms
\cite{HHL2009,CKS2017,LinTong2020,Costa2022}.
The resulting dissipative HHL-type solver implements projected-row
constraints through pumping channels on a single register.
The HHL-type designation refers to the solution-state target;
our construction prepares this state by Lindblad evolution.
The original HHL algorithm uses Hamiltonian simulation, phase
estimation, and conditional eigenvalue inversion \cite{HHL2009}.
The shared pumping law provides a common basis for comparing the
two encodings and their convergence rates.

Let $\mA\in\sC^{n\times n}$, $n\ge2$, be invertible,
let $\|\vb\|=1$, and let $\vx^\star=\mA^{-1}\vb$.
The target state is
\begin{equation*}
    \ket{\vx}
    =\frac{\mA^{-1}\ket{\vb}}{\|\mA^{-1}\vb\|}
    =\frac{\vx^\star}{\|\vx^\star\|}.
\end{equation*}
The penalty construction of
adiabatic quantum linear-system algorithms
\cite{SSO2019,AnLin2022} gives the projected
matrix
\begin{equation*}
    \hat{\mA}=(I-\proj{\vb})\mA.
\end{equation*}
Since $\mA\vx^\star=\vb$ and $\|\vb\|=1$,
we have $\hat{\mA} \vx^\star=0$.
Moreover, invertibility of $\mA$ gives
$\ker\hat{\mA}=\mathrm{span}\{\vx^\star\}$.
Thus, the rows of $\hat{\mA}$ provide homogeneous
constraints whose common solution space is
the target direction.

Let $\mathcal R$ index the nonzero rows of
$\hat{\mA}$. For each $k\in\mathcal R$, define
the normalized row and its projector by
\begin{equation}
    \hat{\va}_k^\dagger
    =
    \frac{\va_k^\dagger-b_k\vb^\dagger \mA}
         {\|\va_k^\dagger-b_k\vb^\dagger \mA\|},
    \qquad
    \hat\Pi_k=\proj{\hat{\va}_k}.
\label{eq:hhl-projectors}
\end{equation}
These projected rows involve the global quantity $\vb^\dagger\mA$.
Applying the pumping law \eqref{eq:rowdiss} to each projected row
on the same $n$-level register with rate $\hat\gamma_k>0$ gives
\begin{equation*}
    \hat\cD_k(\varrho)
    =
    \hat\gamma_k
    \left[
    \bra{\hat{\va}_k}\varrho\ket{\hat{\va}_k}
    \frac{I-\hat\Pi_k}{n-1}
    -
    \frac12
    \bigl(\hat\Pi_k\varrho+\varrho\hat\Pi_k\bigr)
    \right].
\end{equation*}

Each channel transfers population from $\ket{\hat{\va}_k}$ into
$\ker\hat\Pi_k$ at rate $\hat\gamma_k$.
The summed pumping channels determine the single-register dynamics,
with the corresponding dissipation and penalty operators given by
\begin{equation}
    \dot\varrho
    =\sum_{k\in\mathcal R}\hat\cD_k(\varrho),
    \quad
    \hat P_k=-\hat\cD_k,
    \quad
    \hat Q_\gamma
    =\sum_{k\in\mathcal R}\hat\gamma_k\hat\Pi_k.
\label{eq:hhl-me}
\end{equation}
The linear operator $\hat P_k$ acts on the traceless Hermitian
operators on $\sC^n$. By Lemma~\ref{lem:row}, $\proj{\vx}$ is the unique
common dark state of
$\{\hat\cD_k\}_{k\in\mathcal R}$.
Let $\hat\mu$ be the smallest positive eigenvalue
of $\hat Q_\gamma$ and $\hat\lambda$ the smallest
eigenvalue of $\sum_{k\in\mathcal R}\hat P_k$
on this space.

Choosing squared projected-row norms as pumping rates cancels
the normalization
in \eqref{eq:hhl-projectors}:
\begin{equation}
    \hat\gamma_k
    =\|\va_k^\dagger-b_k\vb^\dagger \mA\|^2,
    \quad
    \hat Q_\gamma
    =\mA^\dagger(I-\proj{\vb})\mA.
\label{eq:hhl-norm-rates}
\end{equation}
Under this rate choice,
$\hat\mu\ge\sigma_{\min}(\mA)^2$
\cite{Dalzell2024}.

\begin{proposition}
\label{prop:hhl}
Consider the dissipative HHL-type solver
\eqref{eq:hhl-me}.
The state $\varrho(t)$ converges exponentially
to the unique stationary state $\proj{\vx}$,
satisfying
\begin{equation}
    \|\varrho(t)-\proj{\vx}\|_2
    \le e^{-\hat\lambda t}
    \|\varrho(0)-\proj{\vx}\|_2.
\label{eq:hhl-decay}
\end{equation}
Moreover, the convergence rate $\hat\lambda$ satisfies
\begin{equation}
    \frac{\hat\mu}{n}
    \le\hat\lambda
    \le\frac{\hat\mu}{2}.
\label{eq:hhl-rate}
\end{equation}
With the squared-norm rates \eqref{eq:hhl-norm-rates},
it further satisfies
\begin{equation}
    \hat\lambda
    \ge\frac{\sigma_{\min}(\mA)^2}{n}.
\label{eq:hhl-condition-rate}
\end{equation}
\end{proposition}

The dimension dependence in Proposition~\ref{prop:hhl} reflects
uniform population redistribution across the complementary subspace.
The spectral bounds quantify its contribution to convergence toward
the common dark state.

\subsection{Encoding and Rate Comparison}
\label{sec:hhl-vs}

We compare the two solvers for invertible $\mA\in\sC^{n\times n}$
and $\|\vb\|=1$. The distributed quantum solver uses
$N=m=n$ and register dimension $d=n+1$ at each node.
The two encodings are connected through the
homogenizing coordinate.
With squared row norms as pumping rates,
the penalties in \eqref{eq:tls-objective}
and \eqref{eq:hhl-norm-rates} satisfy,
for every $\vy\in\sC^n$,
\begin{equation}
    \vy^\dagger\hat Q_\gamma \vy
    =
    \min_{\tau\in\sC}\|\mA\vy-\tau \vb\|^2
    =
    \min_{\tau\in\sC}
    (\vy,\tau)^\dagger Q(\vy,\tau).
\label{eq:schur}
\end{equation}
The minimum is attained at $\tau=\vb^\dagger \mA\vy$.
Thus, $\hat Q_\gamma$ is the Schur complement
of $Q$ for the homogenizing coordinate.

The corresponding target states are related
by measurement.
By Theorem~\ref{thm:consistent}, each node of
the distributed quantum solver converges to
$\proj{\xt}$.
A measurement with projectors $I-\proj d$
and $\proj d$ extracts the normalized solution
state from this limiting state with probability
\begin{equation*}
    p_{\rm sol}
    =\frac{\|\vx^\star\|^2}{1+\|\vx^\star\|^2}.
\end{equation*}
Conditioned on the first outcome, the state
in the first $n$ levels is $\proj{\vx}$,
the target of the dissipative HHL-type solver.
The following proposition relates the decay rates of the two
encodings under squared-norm pumping rates.

\begin{proposition}
\label{prop:net-hhl}
Consider both the distributed quantum solver
\eqref{eq:network} and the dissipative HHL-type
solver \eqref{eq:hhl-me}.
With the squared-norm rates, their convergence
rates satisfy
\begin{equation}
    \frac{\sigma_{\min}(\mA)^2}{n(n+1)}
    \le\bar\lambda
    \le\frac{\hat\lambda}{2}.
\label{eq:net-hhl-rate}
\end{equation}
\end{proposition}

The two solvers use the same pumping law but differ in data access
and register organization. The distributed quantum solver encodes
each original row locally and couples separate registers through
neighboring swaps. The dissipative HHL-type solver applies all
projected-row dissipators to one register and requires the global
quantity $\vb^\dagger\mA$. Consequently, changes in the equation
data can be incorporated locally in the distributed construction,
whereas the projected encoding may require broader reconfiguration.
Proposition~\ref{prop:net-hhl} compares the decay rates of these
two pumping constructions under the specified squared-norm rates.
This comparison does not establish a gate-complexity relation to HHL.

\section{Numerical Results}
\label{sec:numerics}

\subsection{Convergence and Solution Recovery}
\label{sec:num-recovery}

This subsection examines state convergence and solution recovery
for consistent equations, followed by least-squares recovery for
inconsistent equations. All random samples in the numerical
experiments are generated with fixed seeds.
The Gaussian examples below use standard real Gaussian entries
for matrices, trial solutions, and noise vectors; consistent
instances use $\vb=\mA\vx^\star$.

\begin{figure}[h]
\centering
\includegraphics[width=0.98\textwidth]{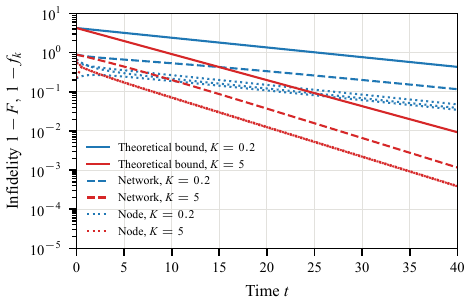}
\vspace{-4pt}
\caption{Unique-solution recovery: bounds (solid), network
infidelity $1-F(t)$ (dashed), and node infidelities
$1-f_k(t)$ (dotted).}
\label{fig:consistent-unique}
\end{figure}

We first consider a unique-solution system on a three-node path with $m=2$,
unit pumping rates, and a random pure product initial state.
Matrix-exponential trajectories in Fig.~\ref{fig:consistent-unique}
show both node and network infidelities decreasing within the
theoretical exponential envelopes. This demonstrates recovery
of the local solution states and the joint product target,
the two convergence conclusions of Theorem~\ref{thm:consistent}.
Increasing the swap gain from $K=0.2$ to $K=5$ accelerates
recovery on the same network.

\begin{figure}[h]
\centering
\includegraphics[width=0.98\textwidth]{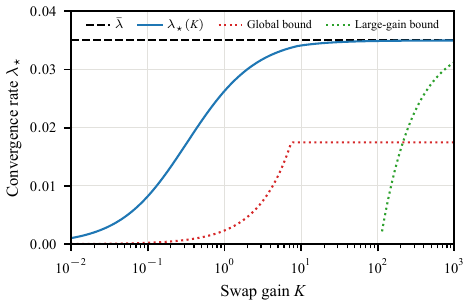}
\caption{Computed rate $\lams(K)$ compared with its limit
$\lamb$ and the global and large-gain lower bounds.}
\label{fig:consistent-rate}
\end{figure}

To examine gain dependence, we fix another consistent instance
with $N=12$ and $m=3$.
The rows occupy fixed nodes of a unit-weight ring with unit
pumping rates.
For $K\in[10^{-2},10^3]$, the computed rate in
Fig.~\ref{fig:consistent-rate} increases toward $\lamb$ and
remains above both theoretical lower bounds. The gain-induced
acceleration and subsequent saturation match the monotonicity
and limiting-rate conclusions of Theorem~\ref{thm:rates}:
larger swap gains bring the rate closer to the limit set by
the local dissipators.

\begin{figure}[h]
\centering
\subfloat[State convergence]{\includegraphics[width=0.49\textwidth]{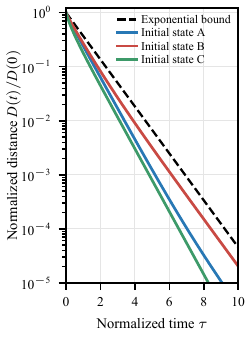}}
\hfill
\subfloat[Solution readout]{\includegraphics[width=0.49\textwidth]{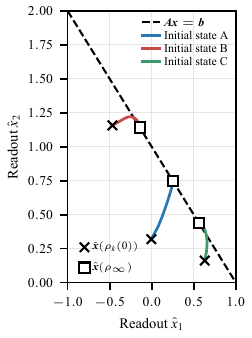}}
\caption{Evolution from three initial states under the same
equation and graph. Dashed lines show the exponential bound in
(a) and the projected solution set in (b); crosses and squares
mark initial and limiting readouts.}
\label{fig:consistent-multiple}
\end{figure}

The multiple-solution example uses a fixed rank-two consistent system
$\mA\in\mathbb R^{4\times3}$ on a unit-weight path
with $K=1.5$ and unit pumping rates.
We consider three initial states, each shared by all nodes,
and compute the trajectories by matrix exponentials.
We measure convergence by the normalized node-tuple distance
$D(t)/D(0)$, where
\begin{equation*}
D(t)=\|\boldsymbol\rho(t)-\boldsymbol\rho_\infty\|_2.
\end{equation*}
Figure~\ref{fig:consistent-multiple}(a) uses normalized time
$\tau=\lams(K)t$, where $t$ is the evolution time.
All three normalized distances remain below $e^{-\tau}$,
showing contraction toward their respective limiting states
at least at the marginal rate $\lams(K)$ guaranteed by
Theorem~\ref{thm:consistent}. In panel~(b), the first two readout
coordinates at node $1$ approach distinct points on the same
projected solution line. These different limiting readouts reflect
the initial-state dependence of $\rho_\infty$ in \eqref{eq:limit}.

Finally, we examine least-squares recovery on a weighted 10-node ring with
$m=3$ and $\vb=\mA\vx^\star+0.6\boldsymbol\xi$.
We use the scaled encoding and squared-row-norm pumping rates,
with $\beta\in[10^{-3},1]$ and $K\in[1,10^5]$.
Linear solves give the node equilibria. We measure the worst-node
relative LS readout error $E_{\rm LS}$, the relative readout bias
$B_{\rm LS}$ at $\rss(\beta)$, the stationary state of the
averaged dynamics \eqref{eq:averaged-dynamics}, and the worst-node
state distance $E_\rho$ from this state:
\begin{equation*}
\begin{aligned}
E_{\rm LS}(K,\beta)
&=\max_k\frac{\|\hat{\vx}^\beta(\rho_k^{K,\beta})-\vx_{\rm LS}\|}
{\|\vx_{\rm LS}\|},\\
B_{\rm LS}(\beta)
&=\frac{\|\hat{\vx}^\beta(\rss(\beta))-\vx_{\rm LS}\|}
{\|\vx_{\rm LS}\|},\\
E_\rho(K,\beta)
&=\max_k\|\rho_k^{K,\beta}-\rss(\beta)\|_2.
\end{aligned}
\end{equation*}

\begin{figure}[h]
\centering
\subfloat[Readout error and bias]{\includegraphics[width=0.49\textwidth]{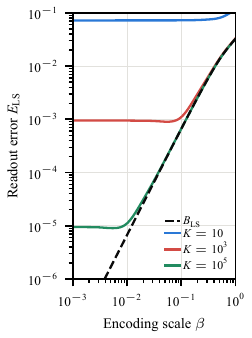}}
\hfill
\subfloat[Large-gain convergence]{\includegraphics[width=0.49\textwidth]{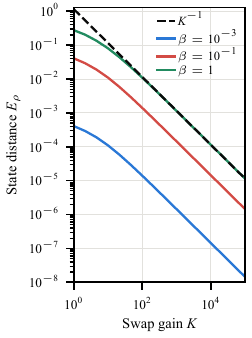}}
\caption{Readout bias and large-gain convergence.
Dashed curves show $B_{\rm LS}$ in (a) and a $K^{-1}$ reference
in (b). Curves in (a) interpolate samples in log--log coordinates.}
\label{fig:network-ls}
\end{figure}

\begin{figure}[h]
\vspace{-4pt}
\centering
\includegraphics[width=0.98\textwidth]{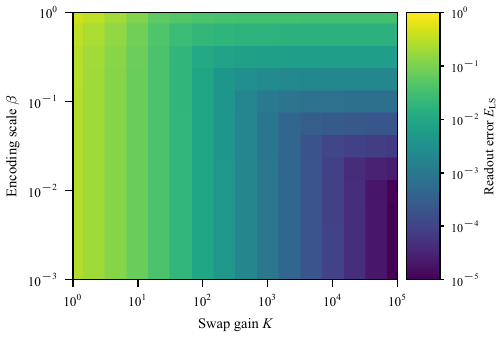}
\caption{LS readout error versus swap gain and encoding scale.}
\label{fig:ls-readout-error}
\vspace{-4pt}
\end{figure}

In Fig.~\ref{fig:network-ls}(a), the readout bias
$B_{\rm LS}$ decreases quadratically with $\beta$, matching the
$O(\beta^2)$ approach to the LS solution in
Theorem~\ref{thm:inconsistent}. Panel~(b) shows $O(K^{-1})$
decay of $E_\rho$ at fixed $\beta$: the node equilibria approach
$\rss(\beta)$ at the order established in
\eqref{eq:inconsistent_perturb}. Figure~\ref{fig:ls-readout-error} shows how swap gain and encoding
scale jointly affect the LS readout error. At finite gain, the error
reaches a plateau as $\beta$ decreases; increasing $K$ lowers this
plateau. The heatmap complements the separate scaling results by
showing the readout accuracy across both parameters.

\vspace{-4pt}
\subsection{Swap-Rate Design}
\label{sec:design-num}

The same 12-row equation and unit pumping rates are used on
path, ring, star, random $3$-regular, and complete
graphs, with row placement fixed. The random $3$-regular (random) graph is connected,
with three neighbors per node.

The total swap-rate budget is $W=\sum_e w_e$.
We use $W_{90}$ to denote the smallest budget at which the marginal
convergence rate reaches $90\%$ of its limiting value $\lamb$.
The superscripts in $W_{90}^{\rm unif}$ and $W_{90}^{\rm opt}$
refer to uniform edge rates and SDP-optimal edge rates, respectively,
with the latter optimized separately at each budget.

\begin{figure}[h]
\centering
\subfloat[Rate deficits]{\includegraphics[width=0.49\textwidth]{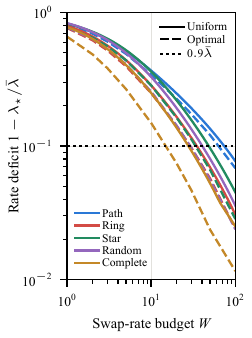}}
\hfill
\subfloat[Edge allocation]{\includegraphics[width=0.49\textwidth]{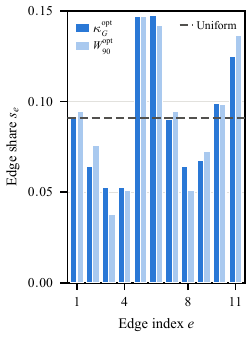}}
\caption{Comparison of uniform and optimized swap rates.
In (a), solid and dashed curves use uniform and rate-optimal
allocations; the dotted line marks $\lams=0.9\lamb$.
In (b), the dashed line indicates uniform allocation.}
\label{fig:network-design}
\end{figure}

Figure~\ref{fig:network-design}(a) shows that SDP-optimal
allocations give smaller rate deficits than uniform allocations
at the same budget. The deficits decrease as $W$ grows across
all five topologies, consistent with the common limiting rate
$\lamb$ and the large-budget convergence established in
Proposition~\ref{prop:design}.
Panel~(b) compares path-edge shares $s_e=w_e/W$ minimizing
$\kappa_G$ and maximizing the rate at $W_{90}^{\rm opt}$;
edge $e$ joins nodes $e$ and $e+1$.

Table~\ref{tab:topology-comparison} reports the large-budget
coefficient $\kappa_G$ and the target-rate budget $W_{90}$;
$\lambda_2$ is computed using uniform shares $s_e=1/|\gE|$.
Here $\kappa_G^{\rm opt}$ is the minimum large-budget coefficient,
whereas $W_{90}^{\rm opt}$ uses allocations maximizing the
finite-budget convergence rate. These are distinct design objectives.
Bold and underlined values indicate the best and second-best
results in each metric column, respectively.
Rate optimization reduces the target budget by 13--46\% across
the five tested topologies.
The topology effect appears in both the rate deficits in
Fig.~\ref{fig:network-design}(a) and the target budgets in
Table~\ref{tab:topology-comparison}. With the same local dissipators,
all five graphs share the limiting rate $\lamb$, but require
different budgets to approach it. The complete graph achieves
the smallest optimized target budget among the tested topologies.
Moreover, the ring requires a smaller uniform-rate target budget
than the star despite having lower algebraic connectivity.
These results show that topology and row placement jointly
affect finite-budget convergence. Their leading large-budget
effect is quantified by the Laplacian-pseudoinverse coefficient
in \eqref{eq:kappa-definition}.

\begin{table}[h]
\centering
\caption{Large-budget coefficients and budgets attaining
$0.9\lamb$ across five topologies.}
\label{tab:topology-comparison}
\footnotesize
\renewcommand{\arraystretch}{1.2}
\setlength{\tabcolsep}{2pt}
\begin{tabular*}{\textwidth}{@{\extracolsep{\fill}}lllllll@{}}
\toprule
Topology
& $|\mathcal E|$
& $\lambda_2$
& $\kappa_G^{\mathrm{unif}}$
& $\kappa_G^{\mathrm{opt}}$
& $W_{90}^{\mathrm{unif}}$
& $W_{90}^{\mathrm{opt}}$ \\
\midrule
Path
& 11 & 0.006 & 0.287 & 0.2526 & 73.6 & 63.9 \\
Ring
& 12 & 0.022 & $\underline{0.107}$
& $\underline{0.0845}$ & $\underline{34.8}$ & 29.1 \\
Star
& 11 & $\underline{0.091}$ & 0.156
& 0.1026 & 48.2 & 33.8 \\
random
& 18 & 0.048 & 0.128
& 0.0854 & 40.4 & $\underline{28.9}$ \\
Complete
& 66 & $\mathbf{0.182}$
& $\mathbf{0.090}$ & $\mathbf{0.0409}$
& $\mathbf{27.6}$ & $\mathbf{15.0}$ \\
\bottomrule
\end{tabular*}
\end{table}

For row placement, we sample $2\times10^4$ assignments on each
path, ring, and star graph.
The sampled placements are ranked by uniform-share $\kappa_G$.
Table~\ref{tab:row-assignment} reports those with the lowest,
median, and highest sampled coefficients.
Bold and underlined values indicate the best and second-best
results within each topology, respectively.
The selected path placements produce a nearly twelvefold range
in $W_{90}^{\rm unif}$, despite having the same equation and
graph. Changing row placement changes the local deviations
$h_k$ and hence their Laplacian-weighted coefficient
$\kappa_G$ in \eqref{eq:kappa-definition}. The budget variation
therefore shows how the interaction between row placement and
topology affects the resources needed to reach the target rate.

\begin{table}[h]
\centering
\caption{Row-placement effects on the large-budget correction
and target-rate budget.}
\label{tab:row-assignment}
\footnotesize
\renewcommand{\arraystretch}{1.2}
\setlength{\tabcolsep}{2pt}
\begin{tabular*}{\textwidth}{@{\extracolsep{\fill}}lllll@{}}
\toprule
Topology
& Placement
& $\kappa_G^{\mathrm{unif}}$
& $W_{90}^{\mathrm{unif}}$
& $W_{90}^{\mathrm{opt}}$ \\
\midrule
\multirow{3}{*}{Path}
& Low
& $\mathbf{0.093}$
& $\mathbf{30.5}$
& $\mathbf{25.0}$ \\
& Median
& $\underline{0.342}$
& $\underline{88.9}$
& $\underline{74.9}$ \\
& High
& 1.328
& 361.5
& 319.9 \\
\addlinespace[2pt]
\multirow{3}{*}{Ring}
& Low
& $\mathbf{0.091}$
& $\mathbf{29.6}$
& $\mathbf{22.6}$ \\
& Median
& $\underline{0.202}$
& $\underline{60.2}$
& $\underline{45.2}$ \\
& High
& 0.447
& 124.4
& 117.5 \\
\addlinespace[2pt]
\multirow{3}{*}{Star}
& Low
& $\mathbf{0.134}$
& $\mathbf{40.6}$
& $\mathbf{29.6}$ \\
& Median
& $\underline{0.176}$
& $\underline{53.8}$
& $\underline{39.0}$ \\
& High
& 0.178
& 54.2
& 39.7 \\
\bottomrule
\end{tabular*}
\end{table}

\vspace{-4pt}
\subsection{Solver Comparison}
\label{sec:num-update}

We compare the solvers after a single-row update to an invertible
$6\times6$ system with $\vb=\one/\sqrt6$.
Row $4$ of $\mA$ is multiplied by $1.45$, with $\vb$ fixed.
Both solvers start at their old equilibria and use squared-row-norm
pumping rates; the distributed solver uses a six-node ring
with $K=5$. Trajectories are computed by matrix exponentials.

\begin{figure}[h]
\centering
\subfloat[Solution recovery]{\includegraphics[width=0.49\textwidth]{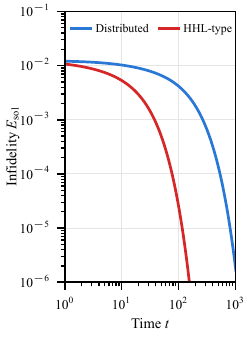}}
\hfill
\subfloat[Dissipator changes]{\includegraphics[width=0.49\textwidth]{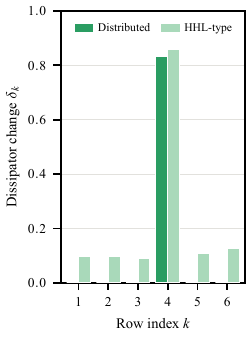}}
\caption{Comparison of the distributed and dissipative HHL-type
solvers after a single-row update.}
\label{fig:solver-update}
\end{figure}

Figure~\ref{fig:solver-update}(a) measures solution recovery using
the solution-state infidelity $E_{\rm sol}$, defined as one minus
the overlap of the normalized output state with the new solution-state
projector. For the distributed quantum solver, we take the largest
infidelity among the nodes' conditional output states; for the
dissipative HHL-type solver, we use its register state.
Under the specified squared-row-norm pumping rates, the computed
generator rates satisfy $\lams(K)\le\lamb\le\hat\lambda/2$.
Here $\lams(K)$ is the finite-gain marginal rate, $\lamb$ its
large-gain limit, and $\hat\lambda$ the dissipative HHL-type rate.
This ordering connects the finite-gain bound in
Theorem~\ref{thm:rates} with the encoding comparison in
Proposition~\ref{prop:net-hhl}.

Figure~\ref{fig:solver-update}(b) measures the relative dissipator
change $\delta_k$ for row $k$,
\begin{equation*}
\delta_k=
\frac{\|\bm D_k^{\rm new}-\bm D_k^{\rm old}\|_2}
{\|\bm D_k^{\rm old}\|_2},
\end{equation*}
where $\bm D_k$ is its column-vectorized matrix representation,
including the pumping rate. Thus, $\delta_k$ quantifies the change
in the complete row pumping dissipator: $\delta_k=0$ indicates
that it is unchanged, while a positive value measures its relative
change in Hilbert--Schmidt norm.
Only one distributed dissipator changes, compared with all six
HHL-type dissipators. Each distributed dissipator uses only its
own augmented row, so changing row $4$ leaves the other local
dissipators intact. The projected rows depend on the shared
quantity $\vb^\dagger\mA$, whose change reconfigures all six
HHL-type dissipators here.

\vspace{-4pt}
\section{Conclusions}

We developed a distributed quantum solver combining row-local
pumping with neighboring swaps.
The exact closed marginal dynamics yield exponential convergence
to common solution-space states for consistent equations and
full-network convergence when the solution is unique.
Scaled encoding and readout recover the least-squares solution
of inconsistent full-column-rank systems through successive
large-gain and small-scale limits.
Rate analysis identifies a graph-independent limit and a
Laplacian-pseudoinverse correction coupling topology and row placement.
These results yield semidefinite edge-rate optimization under a
fixed budget, with sensitivity and allocation optimality conditions.
Applying the same pumping law to projected rows yields a
single-register dissipative HHL-type solver.
Conditional measurement links its solution-state target to the
network encoding, and the comparison characterizes data requirements
and spectral decay rates under squared-row-norm pumping.

\appendix
\section{Proofs of Main Results}
\label{app:proofs}

\subsection{Proof of Theorem~\ref{thm:consistent}}
\begin{proof}
\emph{Marginal convergence.} The reduced operator has quadratic form
\begin{equation}
\begin{aligned}
\langle\boldsymbol Z,M_K(\boldsymbol Z)\rangle
&=K\sum_{\{j,k\}\in\gE}a_{jk}\|Z_j-Z_k\|_2^2\\
&\quad+\sum_k\langle Z_k,P_k(Z_k)\rangle.
\end{aligned}
\label{eq:quadM}
\end{equation}
Connectivity and Lemma~\ref{lem:row} identify its kernel as
\[
\cK_0=\{\one\otimes Z:Z\in\cV_0,\ \Pi_\cN Z\Pi_\cN=Z\}.
\]
Choose the stationary reference $\varrho_0=\Pi_\cN/\dim\cN$.
Projection onto $\cK_0$ retains the traceless part of
$\Pi_\cN\bar\rho\Pi_\cN$; adding back $\varrho_0$ gives
\eqref{eq:limit}.
The spectral theorem on $\cK_0^\perp$ then proves
\eqref{eq:marginal_decay}.

\emph{Fidelity and full-network gap.} For a unique solution,
the union bound for commuting target projectors gives
\[
1-F(t)\le\sum_k(1-f_k(t)).
\]
Since $\|\boldsymbol X(0)\|_2\le\sqrt{2N}$,
Cauchy--Schwarz and \eqref{eq:marginal_decay} give
$\sum_k(1-f_k(t))\le\sqrt N\|\boldsymbol X(t)\|_2$
and hence \eqref{eq:fidelity}.
Each node infidelity is bounded by the network infidelity.
The resulting network trace-distance estimate
\eqref{eq:trace-distance} decays at rate $\lams(K)/2$.
Every traceless operator is a linear combination of differences
of states, so this rate bounds all nonzero generator eigenmodes
and proves the lower bound in \eqref{eq:full-gap-bound}.

For the upper bound, let $\boldsymbol Y$ be a slowest eigenvector
of $M_K$ and set $O=\sum_kY_k^{(k)}$.
Symmetry of $P_k$ on $\cV_0$ implies
$\cD_k^\dagger(Y_k)=-P_k(Y_k)+c_kI$ for some scalar $c_k$.
The swap action and $M_K(\boldsymbol Y)=\lams(K)\boldsymbol Y$
therefore give
\[
\cL^\dagger(O)=-\lams(K)O+cI,\qquad c=\sum_kc_k.
\]
Partial traces show that $O$ is nonzero and traceless.
Thus $O-cI/\lams(K)$ is a nonzero eigenobservable with eigenvalue
$-\lams(K)$, proving the upper bound in \eqref{eq:full-gap-bound}.
\end{proof}

\subsection{Proof of Theorem~\ref{thm:rates}}
\begin{proof}
\emph{Gain dependence.} On $\cK_0^\perp$, the rate is the
minimum of the unit-vector quadratic forms in \eqref{eq:quadM}.
Their affine, nondecreasing dependence on $K$ proves concavity
and monotonicity. A consensus slow mode gives
$\lams(K)\le\lamb$.
Decompose a unit trial tuple as
\[
\boldsymbol Z=\one\otimes Z_{\rm c}+\boldsymbol W,
\qquad \sum_kW_k=0.
\]
Let $a$ and $w$ be the norms of these orthogonal parts.
Lemma~\ref{lem:row} and Cauchy--Schwarz give
\[
\langle\boldsymbol Z,M_K(\boldsymbol Z)\rangle
\ge\frac{\lamb}{2}a^2+(K\lambda_2-\gamma_{\max})w^2.
\]
Since $a^2+w^2=1$ and $\lamb\le\gamma_{\max}$,
the rate is at least $\lamb/2$ at $K=2\gamma_{\max}/\lambda_2$.
Concavity and nonnegativity for smaller gains, and monotonicity
for larger gains, prove \eqref{eq:ratebounds}.

Keeping the cross term instead gives
\[
\langle\boldsymbol Z,M_K(\boldsymbol Z)\rangle
\ge\lamb a^2+K\lambda_2w^2-2\gamma_{\max}aw,
\]
using $\|P_k-\bar P\|\le\gamma_{\max}$.
For $K\lambda_2>\bar\lambda$, completing the square in $w$
and using $a^2+w^2=1$ gives the upper deficit bound in
\eqref{eq:saturation}; $\lams(K)\le\lamb$ gives its lower bound.
This estimate also proves $\lim_{K\to\infty}\lams(K)=\lamb$.

\emph{Limiting-rate bounds.} For $Z\perp\ker\bar P$,
orthogonality gives $\Pi_\cN Z\Pi_\cN=c\Pi_\cN$ for some scalar $c$.
Writing $r=\dim\cN$, the trace-zero condition yields
$rc=-\Tr(\Pi_\cN^\perp Z)$.
Cauchy--Schwarz applied to this identity gives
\begin{equation}
\|Z\|_2^2\le d\Tr(Z^2\Pi_\cN^\perp).
\label{eq:trace-block-estimate}
\end{equation}
Combining \eqref{eq:trace-block-estimate} with
\eqref{eq:Pk-bounds} and the spectral lower bound for $Q$
proves the lower bound in \eqref{eq:limiting-rate-bound}.
For the upper bound, take unit vectors with
$Q\vu=\mu\vu$ and $\vv\in\cN$. Equation~\eqref{eq:Pk} gives
\begin{equation}
Z=\ketbra{\vu}{\vv}+\ketbra{\vv}{\vu},\qquad
\bar P(Z)=\frac{\mu}{2N}Z.
\label{eq:coherence-mode}
\end{equation}
This eigenmode proves the upper bound.

\emph{Large-gain expansion.} Let $D=\mathrm{diag}(P_1,\ldots,P_N)$.
On $\cK_0^\perp$, let $\cC$ be the zero-mode subspace of
the swap operator; its remaining spectrum is separated by
$\lambda_2$. Analytic perturbation theory
\cite[Ch.~II]{Kato1995} gives the effective operator
for the bounded eigenvalue branches:
\[
\begin{aligned}
M_{\rm eff}(K)&=\Pi_\cC D\Pi_\cC\\
&\quad-\frac1K\Pi_\cC D(\mL^+\otimes\id)D\Pi_\cC
+O(K^{-2}).
\end{aligned}
\]
Under the normalized consensus embedding, the leading term
is $\bar P$. Since $\mL^+$ annihilates the common component,
the positive correction compressed to $\bar{\cE}$ has the
quadratic form \eqref{eq:kappa-definition}.
Its largest eigenvalue is $\kappa_G$; the minus sign selects
this eigenvalue for the slowest branch, proving \eqref{eq:kappa}.
\end{proof}

\subsection{Proof of Proposition~\ref{prop:design}}
\begin{proof}
\emph{SDP and budget dependence.} The Rayleigh minimum
\eqref{eq:design-rate} equals the smallest eigenvalue of
the compressed matrix in \eqref{eq:design-sdp}, proving
the SDP formulation.
By \eqref{eq:design-operator}, each trial quadratic form is
affine and nondecreasing in the edge rates.
Their minimum is therefore concave and coordinatewise
nondecreasing. Convex combinations of optimal allocations
at two budgets prove concavity of the optimal value;
increasing an edge rate proves monotonicity.
A consensus test mode bounds the rate by $\lamb$, and
uniform allocations approach this limit by Theorem~\ref{thm:rates}.

\emph{Expansion and convexity.} Taking $K=W$ and edge weights
$\vs$ in \eqref{eq:kappa} gives the expansion.
For a fixed $\bar Z$, collect the zero-sum deviations
from \eqref{eq:kappa-definition} into $\vh$.
Connected support gives the pseudoinverse identity
\[
\mL_{\vs}^+=(\mL_{\vs}+\one\one^\top/N)^{-1}-\one\one^\top/N.
\]
Since $\sum_k h_k=0$, the last term contributes nothing to
$\kappa_G(\vs,\bar Z)$.
The inverse quadratic form is convex because the matrix being
inverted is affine in $\vs$.
Taking the maximum over $\bar Z$ preserves convexity.
\end{proof}

\subsection{Proof of Theorem~\ref{thm:design}}
\begin{proof}
\emph{Edge-rate sensitivity.} A minimizing tuple at $\vw$
is a valid trial tuple at $\vw'$. Substitution into
\eqref{eq:design-rate} gives \eqref{eq:design-supergradient}.

\emph{Allocation optimality.} Under the eigenspace assumption,
the coefficient equals its average over an orthonormal basis
of $\bar{\cE}$. Define the averaged squared potential drop by
\[
\bar g_e=\frac1q\sum_a\|\Phi_{a,j}-\Phi_{a,k}\|_2^2.
\]
Differentiating the inverse quadratic form and applying
the Laplacian energy identity to \eqref{eq:design-potentials} give
\begin{equation}
\frac{\partial\kappa_G}{\partial s_e}=-\frac{\bar g_e}{N},
\qquad \sum_es_e\bar g_e=N\kappa_G(\vs).
\label{eq:allocation-identities}
\end{equation}
Convex first-order optimality requires $\bar g_e$ to be
maximal on used edges and no larger on unused edges.
Sufficiently small feasible redistributions preserve connected
support, so every feasible edge-to-edge transfer can be tested.
Equation~\eqref{eq:allocation-identities} identifies the common
maximum and proves \eqref{eq:equal-drop}.
\end{proof}

\subsection{Proof of Theorem~\ref{thm:inconsistent}}
\begin{proof}
Fix $\beta>0$ and suppress its dependence in the local
operators, rates, and equilibria.

\emph{Unique equilibrium.} The common kernel is trivial, so
Lemma~\ref{lem:row} and \eqref{eq:quadM} make $\bar P$ and
$M_K$ positive definite. Centering the node states gives
\[
\dot{\boldsymbol R}=-M_K(\boldsymbol R)+\vf,
\qquad R_k=\rho_k-I/d,\quad f_k=\cD_k(I/d).
\]
Positive definiteness gives a unique exponentially stable
equilibrium. It corresponds to states because the dynamics
are marginals of a Lindblad evolution.
The averaged dynamics similarly has the unique stationary state
\[
\rss=I/d+\bar P^{-1}(\bar f),
\qquad \bar f=\frac1N\sum_kf_k.
\]

\emph{Large gain.} Set $Y_k=\rho_k^K-\rss$ and decompose
$\boldsymbol Y=\one\otimes\bar Y+\boldsymbol V$, with
$\sum_kV_k=0$.
The equilibrium equation becomes
\[
M_K(\boldsymbol Y)=\ve,\qquad
 e_k=f_k-P_k(\bar P^{-1}(\bar f)).
\]
Stationarity of $\rss$ gives $\sum_ke_k=0$.
Taking the inner product with $\boldsymbol Y$, using nonnegative
local dissipation, and summing over nodes give
\[
K\lambda_2\|\boldsymbol V\|_2^2\le\langle\boldsymbol V,\ve\rangle,
\qquad N\bar P(\bar Y)=-\sum_kP_k(V_k).
\]
Cauchy--Schwarz and $\bar P\succeq\lamb\,\id$ therefore imply
\[
\|\boldsymbol V\|_2\le\frac{\|\ve\|_2}{K\lambda_2},\qquad
\|\bar Y\|_2\le\frac{\gamma_{\max}}{\sqrt N\lamb}
\|\boldsymbol V\|_2.
\]
The triangle inequality therefore gives
\begin{equation}
\max_k\|\rho_k^K-\rss\|_2\le\frac{C}{K},
\label{eq:inconsistent_perturb}
\end{equation}
where $C$ is independent of $K$ for each fixed $\beta$.

\emph{Least-squares readout.} Let $\varrho=\rss(\beta)$,
$\theta_k=\bra{\at_k}\varrho\ket{\at_k}$, and
$w=\sum_k\gamma_k\theta_k$.
Stationarity and squared-row rates give
\begin{equation}
\frac12(Q\varrho+\varrho Q)
=\frac1m\sum_k\gamma_k\theta_k\Pi_k^\perp,
\label{eq:lyap}
\end{equation}
with $Q=\At(\beta)^\dagger\At(\beta)$.
The Rayleigh quotient at $(\beta\vx_{\rm LS},1)$ bounds $\mu$.
For a unit eigenvector $Q\vu=\mu\vu$, testing \eqref{eq:lyap} gives
\[
w=m\mu\bra{\vu}\varrho\ket{\vu}
+\sum_k\gamma_k\theta_k|\bra{\at_k}\vu\rangle|^2.
\]
Since $0\le\theta_k\le1$ and
$\gamma_k|\bra{\at_k}\vu\rangle|^2\le\mu$, we obtain
\begin{equation}
\mu\le\beta^2\|\mA\vx_{\rm LS}-\vb\|^2,\qquad
w\le(m+N)\mu.
\label{eq:ls-mass-bound}
\end{equation}
Thus $w=O(\beta^2)$, as is each $\theta_k$ with $\va_k\ne0$.

Let $\varrho_{xx}$ be the leading $m\times m$ block of $\varrho$
and $\varrho_{dd}=\bra d\varrho\ket d$.
Full column rank and Cauchy--Schwarz give
\[
\sigma_{\min}(\mA)^2(I-\proj d)
\preceq2Q+2\beta^2\|\vb\|^2\proj d.
\]
Taking its expectation and using
$\Tr(Q\varrho)=w$ and \eqref{eq:ls-mass-bound} yields
\begin{equation}
\Tr(\varrho_{xx})=O(\beta^2),\qquad
\varrho_{dd}=1-O(\beta^2).
\label{eq:ls-population-bound}
\end{equation}
The readout is therefore well-defined for small $\beta$.

Writing $\hat{\vx}^\beta=\hat{\vx}^\beta(\varrho)$,
the off-diagonal block of \eqref{eq:lyap} gives
\begin{equation}
\begin{aligned}
\bigl(\mA^\dagger\mA+\beta^2\|\vb\|^2I\bigr)\hat{\vx}^\beta
&=\Bigl(I+\frac{\varrho_{xx}}{\varrho_{dd}}\Bigr)\mA^\dagger\vb\\
&\quad+\frac{2}{m\varrho_{dd}}
\sum_{k:\,\va_k\ne0}\theta_kb_k\va_k.
\end{aligned}
\label{eq:ls-identity}
\end{equation}
Equations~\eqref{eq:ls-mass-bound}--\eqref{eq:ls-population-bound}
make both corrections to $\mA^\dagger\vb$ of order $\beta^2$.
Invertibility of $\mA^\dagger\mA$ then turns
\eqref{eq:ls-identity} into \eqref{eq:ls-limit}.
Combining \eqref{eq:inconsistent_perturb} with readout continuity
proves the successive limit.
\end{proof}

\subsection{Proof of Proposition~\ref{prop:hhl}}
\begin{proof}
\emph{Convergence.} The projected rows span the orthogonal
complement of the solution, so the target is the unique common
dark state. A traceless operator supported on this one-dimensional
space is zero, so Lemma~\ref{lem:row} makes
$\hat P=\sum_{k\in\mathcal R}\hat P_k$ positive definite on the
traceless Hermitian space.
Thus,
\[
X=\varrho-\proj{\vx},\qquad \dot X=-\hat P(X),
\]
and the spectral theorem proves \eqref{eq:hhl-decay}.

\emph{Rate bounds.} Apply \eqref{eq:Pk-bounds} and the block
estimate \eqref{eq:trace-block-estimate} with $d=n$ and
target kernel $\mathrm{span}\{\vx\}$.
The Rayleigh principle gives the lower bound in
\eqref{eq:hhl-rate}.
The coherence construction \eqref{eq:coherence-mode}, applied
to the projected rows and their summed dissipation operator,
gives the eigenvalue $\hat\mu/2$ and hence the upper bound.
The squared-norm estimate for $\hat\mu$ gives
\eqref{eq:hhl-condition-rate}.
\end{proof}

\subsection{Proof of Proposition~\ref{prop:net-hhl}}
\begin{proof}
The penalties in \eqref{eq:tls-objective} and
\eqref{eq:hhl-norm-rates} satisfy
\begin{equation}
\mu\ge\sigma_{\min}(\mA)^2,\qquad \hat\mu\ge\mu.
\label{eq:penalty-comparison}
\end{equation}
For the first inequality, the nonzero eigenvalues of
$Q=\At^\dagger\At$ equal those of
$\mA\mA^\dagger+\vb\vb^\dagger$; apply eigenvalue monotonicity.

For the second, lift a unit vector $\vy\perp\vx^\star$ by
$\vv=(\vy,\vb^\dagger\mA\vy)$.
Its squared distance from the kernel of $Q$ is
\[
\|\Pi_\cN^\perp\vv\|^2
=1+\frac{\|\vx^\star\|^2|\vb^\dagger\mA\vy|^2}
{1+\|\vx^\star\|^2}.
\]
The Schur-complement identity \eqref{eq:schur} and the spectral lower bound for
$Q$ therefore give
\[
\vy^\dagger\hat Q_\gamma\vy=\vv^\dagger Q\vv
\ge\mu\|\Pi_\cN^\perp\vv\|^2\ge\mu.
\]
Minimizing over $\vy$ proves the second inequality in
\eqref{eq:penalty-comparison}.
With $N=m=n$, combining \eqref{eq:penalty-comparison}
with \eqref{eq:limiting-rate-bound} and \eqref{eq:hhl-rate}
gives both bounds in \eqref{eq:net-hhl-rate}.
\end{proof}

\bibliographystyle{agsm}
\bibliography{reference}

\end{document}